\documentclass[conference]{IEEEtran}
\IEEEoverridecommandlockouts
\usepackage{cite}
\usepackage{amsmath,amssymb,amsfonts}
\usepackage{algorithmic}
\usepackage{graphicx}
\usepackage{textcomp}
\usepackage{multirow}
\usepackage{booktabs}
\usepackage{caption}
\usepackage{subcaption}
\usepackage{xcolor}
\usepackage{svg}
\usepackage{url}
\usepackage[all]{nowidow}
\usepackage[flushleft]{threeparttable}

\def\BibTeX{{\rm B\kern-.05em{\sc i\kern-.025em b}\kern-.08em
    T\kern-.1667em\lower.7ex\hbox{E}\kern-.125emX}}
\begin{document}

\title{Supporting Multi-Cloud in Serverless Computing\\
}

\author{
    \IEEEauthorblockN{Haidong Zhao\IEEEauthorrefmark{1}\IEEEauthorrefmark{2}\IEEEauthorrefmark{3}
    \hspace{5pt}Zakaria Benomar\IEEEauthorrefmark{1}
    \hspace{5pt}Tobias Pfandzelter\IEEEauthorrefmark{2}
    \hspace{5pt}Nikolaos Georgantas\IEEEauthorrefmark{1}}
    \IEEEauthorblockA{\IEEEauthorrefmark{1}\it Inria \quad \IEEEauthorrefmark{2}\it Technische Universit\"{a}t Berlin \quad \IEEEauthorrefmark{3}\it Sorbonne Universit\'{e}}

}
\maketitle

\begin{abstract}
Serverless computing is a widely adopted cloud execution model composed of Function-as-a-Service (FaaS) and Backend-as-a-Service (BaaS) offerings. The increased level of abstraction makes vendor lock-in inherent to serverless computing, raising more concerns than previous cloud paradigms. Multi-cloud serverless is a promising emerging approach against vendor lock-in, yet multiple challenges must be overcome to tap its potential. First, we need to be aware of both the performance and cost of each FaaS provider. Second, a multi-cloud architecture needs to be proposed before deploying a multi-cloud workflow. Domain-specific serverless offerings must then be integrated into the multi-cloud architecture to improve performance and/or save costs. Finally, we require workload portability support for serverless multi-cloud.

In this paper, we present a multi-cloud library for cross-serverless offerings. We develop an analysis system to support comparison among public FaaS providers in terms of performance and cost. Moreover, we present how to alleviate data gravity with domain-specific serverless offerings. Finally, we deploy workloads on these architectures to evaluate several public FaaS offerings.
\end{abstract}

\begin{IEEEkeywords}
serverless, multi-cloud, performance and cost, vendor lock-in
\end{IEEEkeywords}

\section{Introduction}
\label{sec:introduction}
Serverless computing is a promising cloud execution model that supports event-driven resource provisioning while eliminating the need for resource management and allocation. This paradigm enables a fine-grained pay-as-you-go billing model~\cite{serverless_view}. Given its widespread adoption, cloud vendors improve and differentiate their serverless offerings, aiming at locking in their clients, as has already been the case with previous cloud computing paradigms~\cite{petcu2011portability}.

Vendor lock-in is exacerbated in serverless computing as public providers only allow customers to use Function-as-a-Service (FaaS) and Backend-as-a-Service (BaaS) offerings~\cite{serverless_view}. More importantly, a public BaaS offering can only natively trigger a FaaS offering from the same provider. Customers are locked into a single provider due to this \emph{tight coupling}. 

Migrating even a simple workload may encounter tricky issues or possibly a dead-end~\cite{yussupov2019facing,hartauer2022cloud,9305905}. 
One solution to mitigate the effect of vendor lock-in is to deploy multi-cloud~\cite{castro2022hybrid}. 
Multi-cloud is not only a technical problem but also an economic and social one. Several factors may propel us to adopt and deploy it.
First, it can increase availability. Second, we cannot make sure that the currently used FaaS provider is suitable for every workload, as FaaS providers typically have different compute-to-memory ratios, resource management policies~\cite{216063}, and pricing schemes. Third, FaaS providers may offer heavy discounts to a company that holds a contract with them. Fourth, FaaS is still in its infancy and lacks geo-distributed deployment. To meet regulatory constraints and improve performance, a company may have to adopt a new vendor that enables local FaaS deployment. As a result, we employ an analysis system that takes into account both performance and cost factors.

A real workload in serverless computing is typically supported by both FaaS and BaaS. To extend a workload across clouds, a well-designed multi-cloud architecture to manage these serverless offerings should be used. 
Besides, domain-specific offerings from small providers may need to be integrated into multi-cloud architectures to make up for what dominant public providers cannot support, helping solve some long-lasting issues such as data gravity~\cite{data_gravity}. 
As more and more data accumulates, it becomes increasingly difficult for data portability since cloud providers charge prohibitively expensive fees for data egress. 
Data gravity also worsens the vendor lock-in in serverless computing.
Consider that using a hardware-accelerated FaaS to retrieve data from another provider's storage service can incur a expensive data egress fee. On the other hand, it is typically free to move data from a storage service to the same provider's FaaS. 
Thus, this locks customers to a single provider and may limit serverless applications to simple workloads.

The Serverless Framework~\cite{serverless_framework} can help in deploying serverless multi-cloud applications. An early effort~\cite{microsoft_multicloud} with it struggled to orchestrate a well-designed multi-cloud architecture, including synchronizing data from multiple storage providers and paying a data egress fee at least twice due to its limited application templates.
Hence, a multi-cloud library would be more practical for designing a multi-cloud architecture with flexible vendor choices and facilitating workload portability.


In summary, this paper\footnote{Our software is in progress and we make it available as open-source at \url{https://github.com/hd-zhao/serverless_multicloud}} makes the following contributions:
\begin{enumerate}
\item We discuss the motivation of multi-cloud deployments as well as the necessity of redesigning a multi-cloud library for serverless computing (Section~\ref{sec:motivation}).

\item We propose a serverless multi-cloud library that supports workload deployment and portability among clouds with ease (Section~\ref{library_design}).
\item We implement the End Analysis System (EAS) to support the performance evaluation and \emph{fine-grained} cost estimation of public FaaS providers (Section~\ref{end_system}).
\item We discuss the design of multi-cloud architectures and present proof-of-concept architectures that include domain-specific BaaS offerings. In addition, we report the benchmarking results of public FaaS providers in these architectures (Section~\ref{cha:evaluation}).

\end{enumerate}

\section{background and Motivation}
\label{sec:motivation}

\subsection{Multi-Cloud Classification}
\label{sec:multicloud}
We investigated multi-cloud deployment in both industry and academia~\cite{chasins2022sky,multicloud-trend}, and classify the multi-cloud in terms of workload distribution: 

\textbf{Each workload only runs on a single cloud.} One classical case is hybrid clouds, which migrate some workloads to public clouds. According to a survey~\cite{status_multicloud}, the hybrid cloud approach dominates the multi-cloud market. Another representation of this classification is running segmented solutions in different clouds. For example, using Google Docs for collaborative writing and Outlook's email service.

\textbf{A workload runs across multiple clouds.} This refers to the workload migration enabled. This deployment requires us to choose shared services among providers. Deploying a workload across multiple clouds can improve availability and help in finding the best FaaS provider for each workload.

\textbf{A workload is divided among multiple clouds.} A pipeline task or a workload can be separated into several stages. At each stage, we can choose an optimal cloud offering instead of all stages in a single cloud. Researchers from Berkeley proposed a transparent cloud targeting this objective~\cite{stoica2021cloud,chasins2022sky}. A FaaS offering that collaborates with another provider's BaaS offering is also regarded as this multi-cloud category.

\subsection{Benefits of Multi-Cloud Deployment}
\label{sec:multicloud_benefit}

\begin{table*}[t]
\setlength\tabcolsep{2.5pt}
\begin{center}
\begin{threeparttable}
\caption{Pricing scheme of public FaaS providers}
\begin{tabular}{ccccccccc}
\toprule
\multirow{2}{*}[-0.4em]{\textbf{Provider}}&
\multicolumn{2}{c}{\textbf{Duration Fee (GB-second)}}& \multicolumn{2}{c}{\textbf{Invocation Fee (million)}}&
\multicolumn{2}{c}{\textbf{Data Egress Fee (GB / Month)}}&
\multicolumn{2}{c}{\textbf{Ephemeral Storage Fee (GB-second)}} \\
\cmidrule(lr){2-3} \cmidrule(lr){4-5} \cmidrule(lr){6-7} \cmidrule(lr){8-9}
&\textit{Free quotas}&\textit{Beyond}&
\textit{Free quotas}&\textit{Beyond}&
\textit{Free quotas}&\textit{Data Egress}&
\textit{Free quotas}&\textit{Beyond}\\
\midrule
AWS Lambda\tnote{a}~\cite{lambda_price} & 400,000 & \$\(1.67\times 10^{-5}\)&1&\$0.20&100 GB&\$0.09& 512 MB &\$\(3.58\times{10^{-8}}\) \\
Google Cloud Functions\tnote{b}~\cite{gcf_price} &400,000&\$\(2.5\times 10^{-6}\)&2&\$0.40&5 GB &\$0.12&
\multicolumn{2}{c}{Integrated with memory}\\
Alibaba Function Compute~\cite{alibaba_price}&400,000&\$\(1.67\times 10^{-5}\)&1&\$0.20&
\multicolumn{2}{c}{\$0.07 in Frankfurt and London}&512 MB&NAS File System\\
\bottomrule

\end{tabular}
\begin{tablenotes}
    \small
    \item [a]x86 pricing scheme with tiered data egress pricing: \$0.09 in the first 10 TB, \$0.085 in the next 40 TB, and \$0.07 in the following 100 TB.
    \item [b]Tier 1 location pricing. GCF also charges for CPU time
    
\end{tablenotes}
\label{cost_table}
\end{threeparttable}
\end{center}
\end{table*}
\textbf{Mitigate the risks of data gravity and being locked into a FaaS provider.} Ingress data is typically free of charge by cloud providers. However, when data flees from cloud providers to the Internet, providers usually charge a data egress fee, as shown in Table~\ref{cost_table}. Consider if we want to retrieve 10 TB of data from a storage site in London and the storage service is provided by Alibaba, we need to pay \$143.36 to save a month and \$716.8 to move out, 5 times the price difference. With Amazon S3~\cite{aws_s3}, \$239.76 for saving a month and \$891 for moving out. Even if moving out data with high bandwidth just takes minute-level time, it is much more costly than saving the same amount of data for a month.

This is data gravity. Customers tend to store their data with one provider for performance considerations. Therefore, with data growing, it is more unlikely to move data out with the prohibitively expensive data egress fee. Moreover, FaaS is a data-shipping architecture~\cite{DBLP:conf/cidr/HellersteinFGSS19}. Data is unidirectionally moved to the FaaS. Outbound data from an object storage site to the same provider's FaaS is normally free, but to the Internet is costly, especially for large-scale data. This worsens the lock-in issue and precludes customers from using other domain-specific FaaS, such as FaaS with hardware acceleration. In light of this, we may need to employ some low or free egress fee storage services to the multi-cloud architecture.

\textbf{Increase availability, but what about performance enhancement and cost savings.} Compared with traditional compute instances, deploying multi-cloud with FaaS offerings to increase availability does not suffer from being double charged because of its pay-as-you-go billing model~\cite{altwaijiry2021cloud,serving_dnns_like_cloudwork}. Public FaaS providers have different memory-to-compute, resource management policies~\cite{216063}, pricing schemes, etc. Therefore, one serverless provider may perform well for one workload but may be clumsy in another. Thus, it is necessary to evaluate which FaaS provider is most suitable for each workload. Moreover, public FaaS providers usually have deployments in metropolises but with sporadic deployments in other regions, especially developing countries. To adopt serverless computing, a company in this greenfield may use a new provider's FaaS for its local deployment. A FaaS cooperating with another provider's BaaS offering is also a kind of multi-cloud, as described in Section~\ref{sec:multicloud}.

\textbf{Mitigate the risk of an application being locked to a serverless provider.} Due to the lack of a standard and proprietary serverless offerings among providers, we may find it challenging or impossible to migrate an application~\cite{hartauer2022cloud,yussupov2019facing}. Multi-cloud deployment suggests choosing common services in order to preclude being locked into a vendor. However, a company is also able to reserve some proprietary services after risk assessment.

\subsection{Redesign Multi-Cloud Library}\label{sec:library}
Traditional multi-cloud libraries help in accessing general services with a common interface. In serverless, a multi-cloud library should only focus on FaaS and BaaS offerings. Besides, to tap the full potential of multi-cloud, the library should meet the following requirements:

\textbf{Versatile and Scalable.} In addition to supporting BaaS offerings from well-known public clouds, the serverless multi-cloud library should take into account some nimble providers who may offer domain-specific services.

\textbf{Conversion-enabled and one-for-all.} One is able to use the multi-cloud library to access any BaaS provider's services with ease.
Besides, we also account for a situation where a company and its engineers are integrated and trained by a cloud provider, respectively.
If this company plans to port some workloads to other providers, the multi-cloud library should also provide the conversion from one provider's software development kits (SDKs) to that of another one.
This allows engineers to access different providers' serverless offerings with their accustomed SDKs.

\textbf{Lightweight.} Using too many third-party libraries that the function instance runtime does not provide initially may run out of memory and worsen the "cold start" issue since an instance should spend more time preparing for the runtime (i.e., downloading code and third-party libraries from persistent storage).

\section{Design and Implementation}
\begin{figure}[t]
\centerline{\includegraphics[width=\columnwidth]{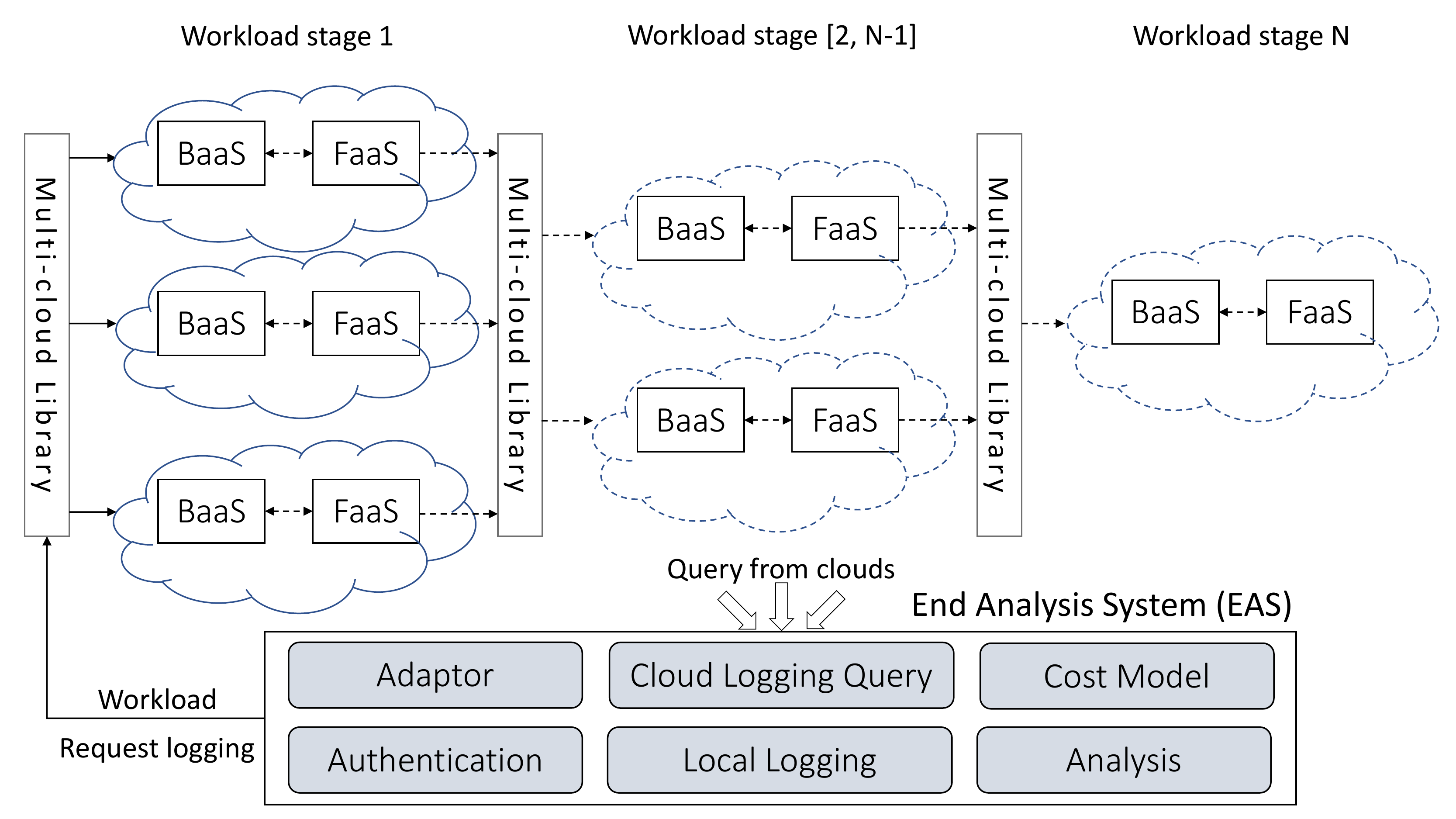}}
\caption{System design overview}
\label{arch}
\end{figure}

Fig.~\ref{arch} depicts the overview of the system design, which is composed of three parts: the workload deployed in the multi-cloud architecture; the multi-cloud library, which enables the utilization of FaaS and BaaS offerings from different vendors with ease; and the End Analysis System, which evaluates the performance and cost of the FaaS providers.

\subsection{System Design Overview}\label{multicloud-arch}
As described in Section~\ref{sec:multicloud_benefit}, a workload may be divided into several stages. Each workload/task stage may work on the most optimized cloud offering. 
Consider that if a workload stage is to store infrequently used data, it may opt for a storage service with the lowest fee for long-term data storage. On the other hand, if this workload stage is for computing, hardware acceleration FaaS may be employed on it. As a result, a workload may be served by more than one cloud provider. 

we design and implement the End Analysis System (EAS) and a serverless multi-cloud library, respectively. To achieve a more accurate workload performance measurement, we may need to log the request that will trigger a serverless offering before sending it. EAS can also be used to evaluate the FaaS providers' performance and cost in a real-world deployment workload, albeit some performance metrics will not be supported. The Multi-cloud library enables us to use common interfaces or familiar SDKs to manage serverless offerings from different providers.

\subsection{The Design of the Multi-Cloud Library} \label{library_design}

One obstacle to enabling workload portability across multiple clouds exists at the API level, where developers may need to learn new SDKs or APIs of new FaaS and/or BaaS offerings from several cloud providers. Thus, we suppose that the multi-cloud library could be the best approach to help developers get familiar with new services.

First, we designed a common interface, acting as an abstraction layer on top of the proprietary APIs or SDKs of cloud providers. In doing so, we make multi-cloud workload deployment easier by allowing developers to use a common interface to access the cloud services that the current multi-cloud library supports. However, considering that if we are currently using a cloud service and want to extend our workloads to multi-cloud deployment, at this moment, we would hope not to learn too many new things, i.e., new providers' SDKs. In view of this, we consider that it is necessary to implement SDK-conversion. To allow this, we further developed provider-like SDKs on top of the proprietary SDKs and APIs for direct conversion, e.g., a mocked AWS S3-like library module that can access other providers' object storage services while maintaining the semantics and syntax of the proprietary SDKs of AWS S3. By doing this, we can access the services of other providers using our familiar SDKs.

\subsection{End Analysis System (EAS)} \label{end_system}
We developed the EAS to benchmark which FaaS provider performs best for each workload in terms of performance and cost. The EAS is composed of the following components:

\textbf{Authentication.}
To gain access to FaaS and BaaS offerings of different providers, the keys or passwords are saved in configuration files. As a request needs to be authenticated in order to access public cloud services, the configuration files are integrated with the multi-cloud library.

\textbf{Adapter.} 
One objective of multi-cloud is to select the best FaaS provider for each workload. FaaS triggered requests are diverse, as a function instance can be triggered via direct HTTP invocation or indirect operations in BaaS offerings. The adapter helps in managing these trigger requests and switching between benchmarked scenarios with ease.

\textbf{Cloud's Logging Query.}
Serverless providers allow developers to debug their programs by reading logs that contain metrics and state information about serverless functions. Cloud logging is typically asynchronous. We query cloud logs from FaaS providers and save them to EAS for further analysis.

\textbf{Local Logging.}
Local logging is important. First, the cloud logs only tell us when a function instance starts and finishes. Second, we may need to evaluate the response time of the request. Moreover, we may need to log the status of running function instances.

\textbf{Cost Model.}
Table~\ref{cost_table} summarizes the pricing schemes of some public serverless providers we use for evaluation in this paper. The duration fee, request fee, and network traffic fee are the three main components of the cost of FaaS offerings. This cost table of FaaS providers is summarized at the time of writing. How to update it as time goes by needs to be discussed. We propose to publish this data on the shared database and maintain the updates on price changes with the help of the community and beneficiaries.

\textbf{Analysis.}
We perform analysis based on the cloud and local logs on two dimensions: performance and cost. For the performance, we focus on the billing duration and latency (request serving latency, response time, etc.). For the cost, we support fine-grained estimation, including the duration fee, the invocation fee, the data egress fee, and perhaps the disk fee.

\section{Evaluation}
\label{cha:evaluation}
\subsection{Proof-of-Concept Multi-Cloud Architectures}\label{pfarch}
As discussed in Section~\ref{sec:multicloud_benefit}, we may also benefit the domain-specific services from some small providers. We propose two proof-of-concept multi-cloud architectures, in which we employ Cloudflare R2, a zero data egress fee storage service.

\subsubsection{Pipeline-Enabled Multi-Cloud Architecture}\label{pfarch1}

\begin{figure}
\centerline{\includegraphics[width=\columnwidth]{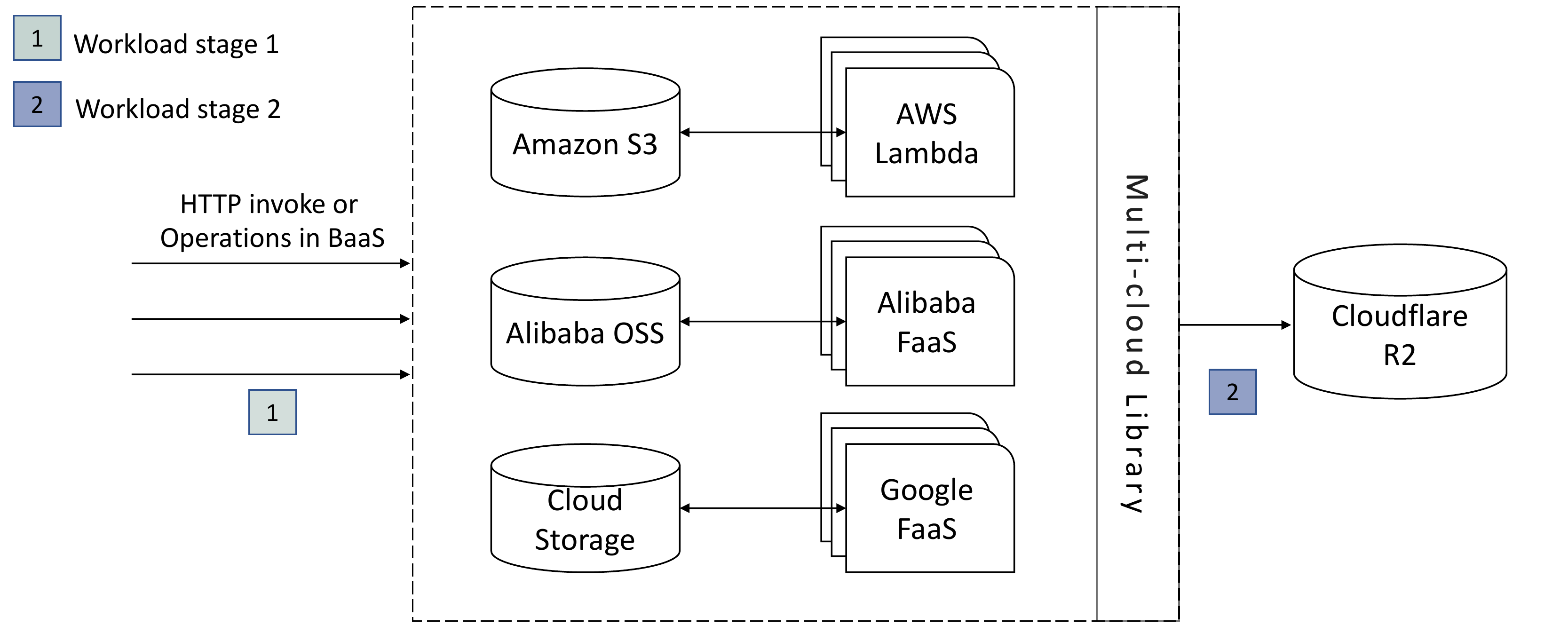}}
\caption{Pipeline-enabled multi-cloud architecture}
\label{arch1}
\end{figure}

Fig.~\ref{arch1} depicts a two-stage workload architecture.
This architecture enables pipeline processing, as we let a FaaS offering be triggered by a BaaS offering from the same provider.

\emph{Optimization 1: }\textit{Avoid data distribution by saving data at one storage site.} The Serverless Framework with limited application templates does not support flexible vendor-choice multi-cloud architectures and may lead to at least double the data egress fee as the data needs to be synchronized across storage providers~\cite{microsoft_multicloud}. What we need to do is to put processed data from FaaS providers into one storage service.

\emph{Optimization 2: }\textit{Leverage the zero egress fee storage service.} In Section~\ref{sec:multicloud_benefit}, we have discussed the prohibitively expensive data egress fee and how the data gravity issue has limited serverless applications. In light of this, we employ Cloudflare R2 to store data in order to save money on network traffic fees and avoid data gravity.

\subsubsection{Pipeline-Disabled Multi-Cloud Architecture}
\label{pfarch2}

\begin{figure}
\centerline{\includegraphics[width=\columnwidth]{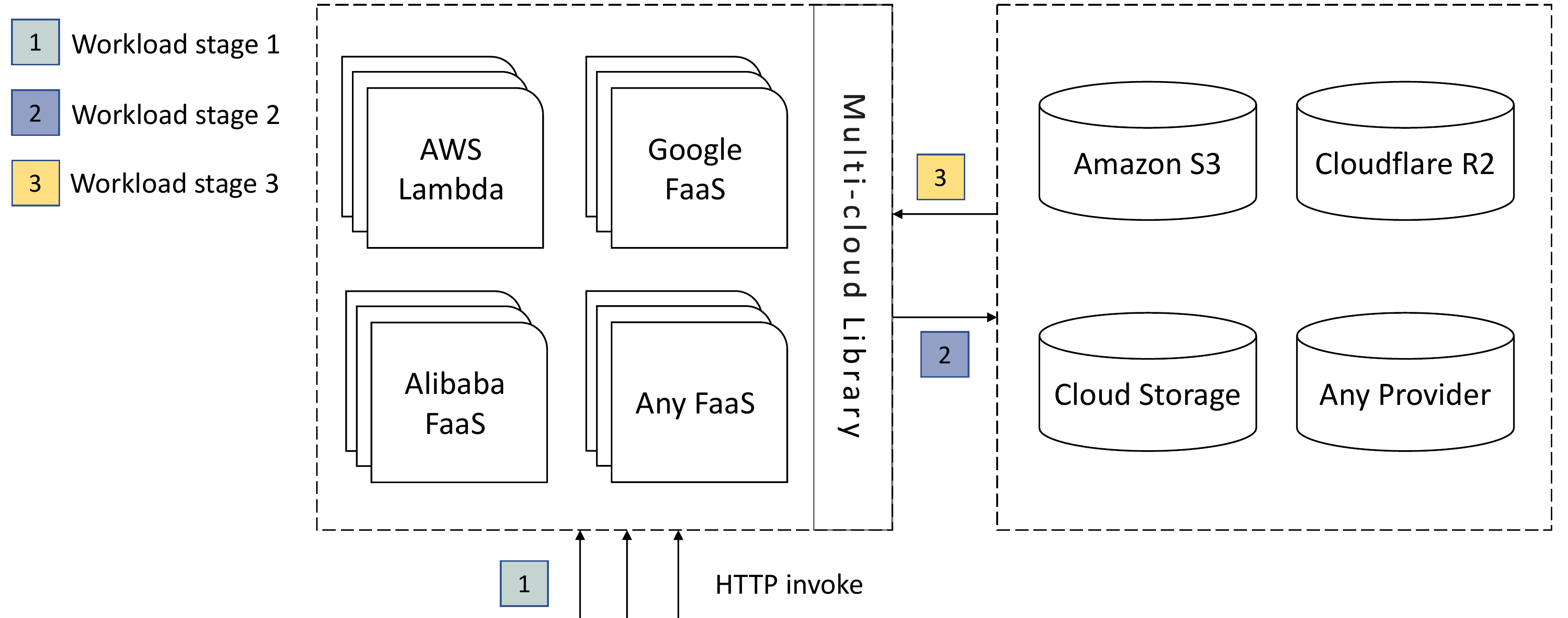}}
\caption{Pipeline-disabled multi-cloud architecture}
\label{arch2}
\end{figure}

Fig.~\ref{arch2} depicts a multi-stage workload architecture.
In this architecture, we further break the boundary between FaaS and BaaS offerings and do not require that they come from the same provider. Therefore, this architecture does not support pipeline processing anymore, as a public provider does not allow its BaaS offering to trigger a FaaS offering from another provider. In the short term, there are no workarounds for it without performance degradation and cost increase.

This multi-cloud architecture may appear in several scenarios. 
For example, a company would like to port some workloads to FaaS but may find that its cooperating cloud vendor does not support local FaaS deployment.
As a result, in order to satisfy its performance and regulatory requirements, this company may need to look for a new cloud provider.
Most public FaaS providers impose time-out limits and rigid compute-to-memory ratios on function instances. Besides, they only support general-purpose hardware.
Nimble FaaS providers, which want to attract customers and differentiate their products, may offer solutions in light of this situation. This would attract more customers to migrate more diverse workloads to it.

\subsubsection{Data Portability}
\label{data_portability}

As discussed in Section~\ref{sec:multicloud_benefit}, data gravity is the real threat, especially in serverless computing, hindering data portability. Cloud providers do not impede data movement but charge a prohibitively expensive data egress fee. As a result, cloud customers may find that they are progressively trapped in data lock-in. This motivates us to employ a zero egress fee storage service in the architecture as shown above. How to migrate data from traditional storage services to this zero egress fee storage service needs to be discussed nevertheless. In an optimal situation, new processed data can be stored in a zero-egress fee storage service first, and then customers can retrieve it from it. This avoids double billing. Otherwise, it should plan how to move redundant data out based on the company's own situation and needs.

\subsection{Benchmark}
Serverless benchmark workloads can be classified into micro-benchmark or application-oriented benchmarks~\cite{FunctionBench,grambow2021befaas,copik2021sebs}. To obtain a more practical evaluation of FaaS providers, we focus on application benchmarks.

\subsubsection{Pipeline-Enabled Multi-Cloud Architecture Benchmark}
In this benchmarking, we stressed the FaaS platforms to test the auto-scaling capability of their providers. 
We sent burst requests to the FaaS platforms, reaching their maximum currency limit.
We set each function instance's timeout to 20s and memory allocation to 1 GB.

\textbf{Image Processing.} 
Specifically, we chose the image resizing application. The function instance downloads the picture from a bucket to disk space before image processing. We chose a data size of 5MB to ensure that the ephemeral storage can handle 50 pictures in a batch (maximum concurrency limit). We use \emph{PILLOW}~\cite{pillow} library to process images.

\subsubsection{Pipeline-Disabled Multi-Cloud Architecture Benchmark}
This architecture may employ the low or zero data egress fee storage service to support domain-specific FaaS, thus supporting data-intensive or computer-intensive serverless workloads.

\textbf{Machine Learning Training.} 
A bulk of papers have discussed how to use FaaS for machine learning jobs~\cite{jiang2021towards,carreira2018case,sanchez2021experience}. There are some positive results show that using FaaS to train some machine tasks performs better, but with more cost for the moment. We trained a machine learning model based on a 100 MB truncated dataset collected from Amazon Fine Foods reviews. We applied TF-IDF (term frequency-inverse document frequency) to evaluate the importance a word has to a document among a set of documents and logistic regression model for classification ,using the machine learning library \emph{scikit-learn}~\cite{scikit-learn}. After cleaning the dataset, we removed the 100 most infrequently used words in text feature extraction and limited the maximum number of iterations to 1000 in the logistic regression model.

\begin{table}[tbp]
\begin{center}
\begin{threeparttable}
\caption{The cost table of experiments}
\begin{tabular}{llll}
\toprule
&\multicolumn{3}{c}{\textbf{Image Processing}}\\
\cmidrule(lr){2-4}
&\textit{AWS}&\textit{Google}&\textit{Alibaba}\\
\midrule
avg. Billing Duration&1937 ms&2421.5 ms &2380.7 ms \\
avg. Egress Data Size&482 KB&482 KB&482 KB\\
99\textsuperscript{th} percentile Latency&5388 ms&12254 ms&9496 ms\\
1M req. Request Fee&\$ 0.2&\$ 0.4&\$ 0.2\\
1M req. Network Fee&\$ 43.38&\$ 57.84 &\$ 33.74 \\
1M req. Duration Fee\tnote{a}&\$ 32.33&\$ 39.95&\$ 39.70\\
1M req. Cost&\$ 75.91&\$ 98.19&\$ 73.64\\ 
1M req. Cost (Free-tier)&\$ 60.05&\$ 94.19&\$ 66.77\\
\bottomrule
\end{tabular}
\begin{tablenotes}
    \small
    \item [a]We merge GCF's CPU charges into duration fee. 

\end{tablenotes}
\label{benchmarking_table}
\end{threeparttable}
\end{center}
\end{table}

\subsection{Experimental Setup}
We deployed the End Analysis System on Azure Standard D8s v3, with 8 vCPUs, 32 GiB of memory, and hosted in Paris. We deployed all our FaaS and BaaS offerings in London.

\subsection{Pipeline-Enabled Multi-Cloud Architecture Evaluation}

\begin{figure}
\centering
\begin{subfigure}{.5\linewidth}
  \centering
  \includegraphics[width=\linewidth]{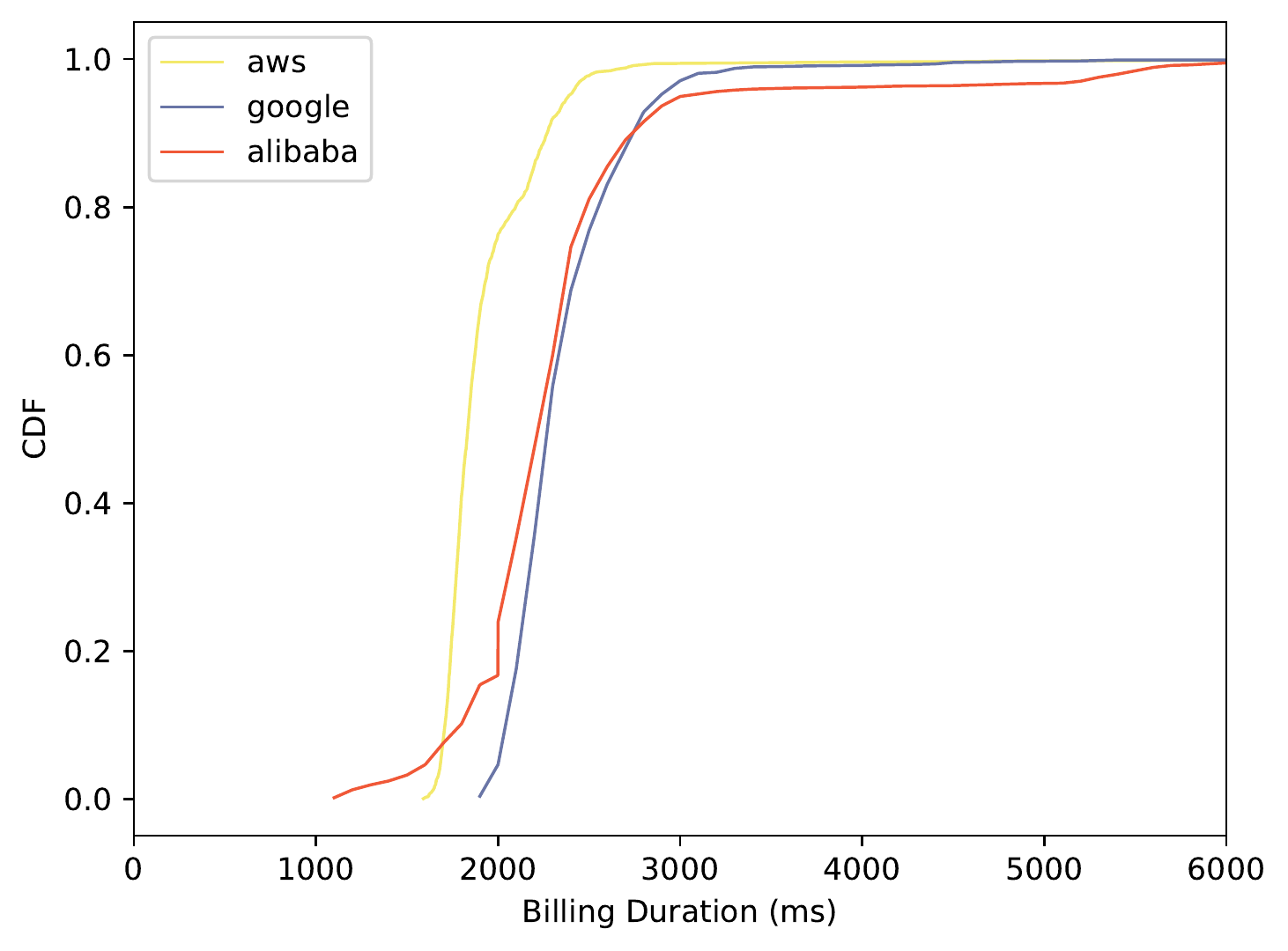}
  \caption{}
  \label{image_processing_a}
\end{subfigure}%
\begin{subfigure}{.5\linewidth}
  \centering
  \includegraphics[width=\linewidth]{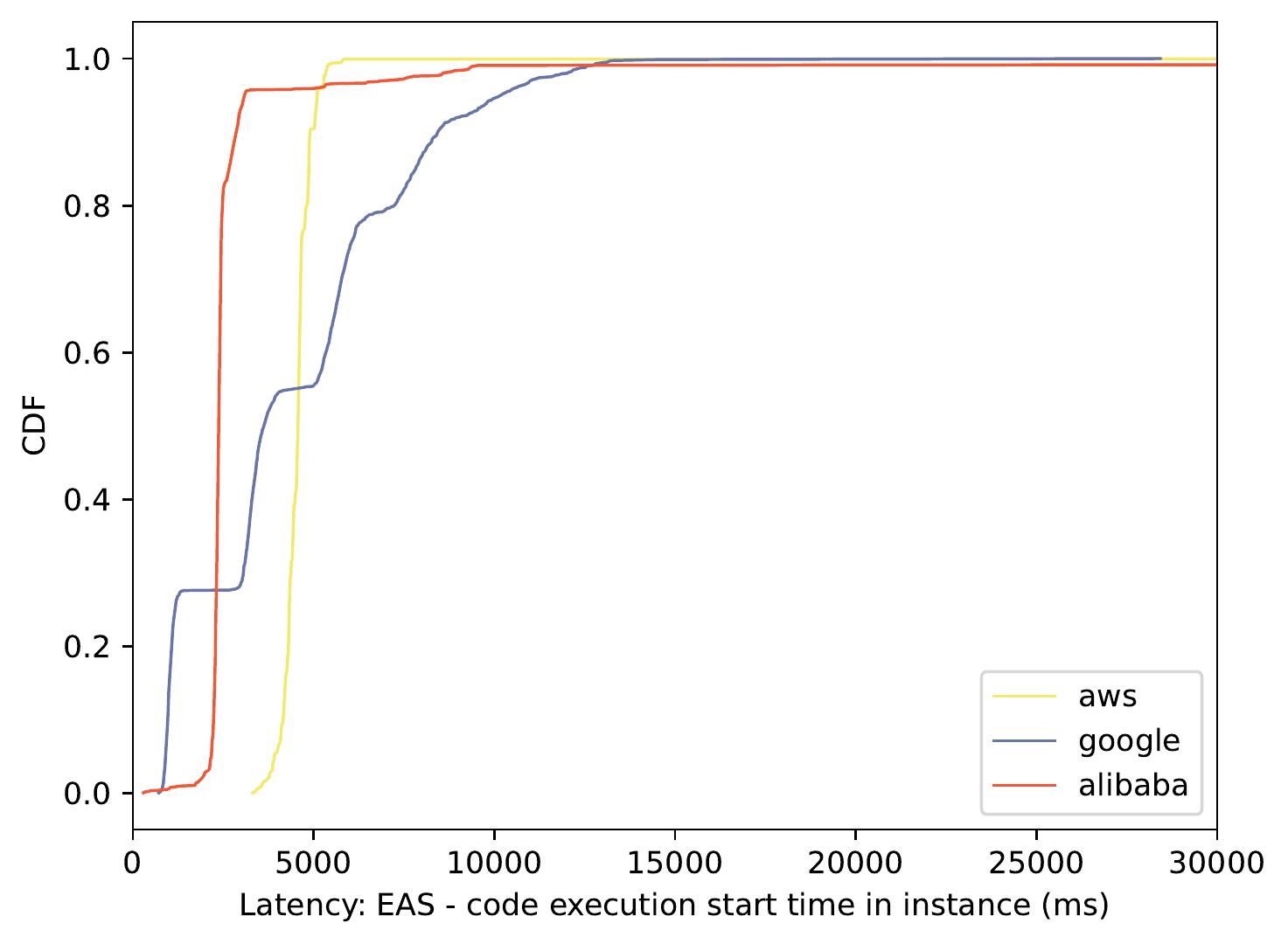}
  \caption{}
  \label{image_processing_b}
\end{subfigure}
\caption{Image Processing}
\label{image_processing}
\end{figure}

\textbf{Image Processing Workload Analysis.}
 Fig.~\ref{image_processing} show the cumulative distribution of evaluated FaaS providers' billing duration and the end-to-end delay, i.e., the elapsed time when a request is handled by a function instance.

\textit{Performance comparison:} As Fig.~\ref{image_processing} and Table~\ref{benchmarking_table} present, AWS Lambda outperforms its counterparts in terms of the lowest average billing duration and 99\textsuperscript{th} percentile latency. Although we may find that Alibaba's most requests have the lowest latency for being served. However, this serving latency is not as stable as AWS Lambda as a certain proportion of requests are served late. As the pressure comes, AWS Lambda may struggle to handle requests as the tail latency of AWS Lambda exceeds 80s. According to the cloud log, the requests that lead to tail latency experience timeout and appear in two scenarios: after the first time-out, the request experiences a long period to be served again; or the first serving time for a request is very late. These scenarios occur when the request is throttled due to a lack of instances to handle.

\textit{Cost comparison:} Table~\ref{benchmarking_table} shows that the overall fee for this workload is dominated by the network fee and the duration fee. AWS Lambda and Alibaba's Function Compute are both economical choices.

In conclusion, AWS Lambda is the best FaaS provider for this workload since it offers the best performance and (one of) the lowest overall costs. Besides, we cannot consider the duration fee as the only main expense for FaaS because the results show that the data egress fee may account for a significant portion of the overall cost.

\subsection{Pipeline-Disabled Multi-Cloud Architecture Evaluation}

\begin{table}
\begin{center}
\begin{threeparttable}
\caption{The FaaS configuration and cost table}
\begin{tabular}{llll}
\toprule
&\multicolumn{3}{c}{\textbf{Machine Learning Training}}\\
\cmidrule(lr){2-4}
&\textit{AWS}&\textit{Google}&\textit{Alibaba}\\
\midrule
Data Size&100 MB & 100 MB & 100 MB\\
vCPU&1.16&1&1.33\\
Memory Allocation&2 GB &2 GB&2 GB\\
Memory Usuage&\(\approx\)700 MB &\(\approx\)675 MB&\(\approx\)720 MB\\
avg. Billing Duration&133 s&159 s &138 s\\
1K req. Duration Fee\tnote{*}&\$ 4.43&\$ 4.64&\$ 4.63\\

\bottomrule

\end{tabular}

\label{ml_table}
\end{threeparttable}
\end{center}
\end{table}

\textbf{ML Training Workload Analysis.}
Table~\ref{ml_table} shows each FaaS provider's compute power when the memory allocation is 2 GB. Even if we chose a small dataset (100 MB) for training traditional ML algorithms, it would take more than 2 minutes to converge. The time-out duration is short in FaaS, which means deploying deep learning algorithms on it will be ineffective for a large training dataset as continued training is needed after a timeout. Besides, we notice the imbalance between memory allocation and usage. We need to allocate more memory to the FaaS in order to gain powerful computing capabilities. However, the computing power increases with the allocated memory for most public FaaS providers. In particular, when training a model with a relatively small dataset, many memory resources will be wasted. We suppose that there would be an opportunity for the FaaS market to support more flexible configurations: GPU enabled (which AFC has enabled), flexible compute-to-memory ratio, etc.

\section{Related Work}
We are motivated to deploy serverless multi-cloud to mitigate the effects of vendor lock-in as migrating an even simplified workload may encounter tricky issues or dead-ends~\cite{yussupov2019facing,hartauer2022cloud}. Lithops is a multi-cloud serverless computing framework for big data analytics and parallel jobs, supporting running the same code across multiple clouds~\cite{Lithops}. Baarzi et al.~proposed developing a latency-aware virtual service provider to schedule requests to public FaaS and proved that multi-cloud provides more stable performance when failure comes~\cite{vsp}. We focus on selecting best FaaS provider for general cases. Some works extended FaaS to on-premises environments and designed scheduling algorithms in the edge-cloud continuum~\cite{shedule2}. Jindal et al.~designed schedule algorithms in a heterogeneous FaaS deployment environment~\cite{shedule1}. Real-world workloads are typically supported by both FaaS and BaaS offerings. This paper also considers the performance of FaaS when it is triggered by BaaS offerings. We are also inspired by Sky Computing, which envisioned a transparent cloud on top of cloud providers to provide best-of-breed service in terms of performance and cost for each workload~\cite{stoica2021cloud,chasins2022sky}.

\section{conclusion}
In this paper, we discuss how to benefit from multi-cloud deployment in addition to mitigating the effect of vendor lock-in. We present two proof-of-concept multi-cloud architectures that aim to avoid some common problems such as data distribution and data gravity. Besides, we developed the End Analysis System to quantitatively evaluate the performance and cost of FaaS providers, assisting customers in selecting the best FaaS provider for their workloads. To facilitate multi-cloud deployment, we designed the first version of a multi-cloud library that allows developers to access resources provisioned by different clouds via a common interface or their familiar SDKs.

\section*{Acknowledgement}
We thank our anonymous reviewers for their valuable feedback that improved this work.

\bibliographystyle{IEEEtran}
\typeout{}
\bibliography{main}

\begin{thebibliography}{10}
\providecommand{\url}[1]{#1}
\csname url@samestyle\endcsname
\providecommand{\newblock}{\relax}
\providecommand{\bibinfo}[2]{#2}
\providecommand{\BIBentrySTDinterwordspacing}{\spaceskip=0pt\relax}
\providecommand{\BIBentryALTinterwordstretchfactor}{4}
\providecommand{\BIBentryALTinterwordspacing}{\spaceskip=\fontdimen2\font plus
\BIBentryALTinterwordstretchfactor\fontdimen3\font minus
  \fontdimen4\font\relax}
\providecommand{\BIBforeignlanguage}[2]{{%
\expandafter\ifx\csname l@#1\endcsname\relax
\typeout{** WARNING: IEEEtran.bst: No hyphenation pattern has been}%
\typeout{** loaded for the language `#1'. Using the pattern for}%
\typeout{** the default language instead.}%
\else
\language=\csname l@#1\endcsname
\fi
#2}}
\providecommand{\BIBdecl}{\relax}
\BIBdecl

\bibitem{serverless_view}
E.~Jonas, J.~Schleier-Smith, V.~Sreekanti, C.-C. Tsai, A.~Khandelwal, Q.~Pu,
  V.~Shankar, J.~Carreira, K.~Krauth, N.~Yadwadkar \emph{et~al.}, ``Cloud
  programming simplified: A berkeley view on serverless computing,''
  \emph{arXiv:1902.03383}, 2019.

\bibitem{petcu2011portability}
D.~Petcu, ``Portability and interoperability between clouds: challenges and
  case study,'' in \emph{European conf. a service-based internet}, 2011, pp.
  62--74.

\bibitem{yussupov2019facing}
V.~Yussupov, U.~Breitenb{\"u}cher, F.~Leymann, and C.~M{\"u}ller, ``Facing the
  unplanned migration of serverless applications: A study on portability
  problems, solutions, and dead ends,'' in \emph{Proc. 12th IEEE/ACM Int. Conf.
  Utility and Cloud Computing}, 2019, pp. 273--283.

\bibitem{hartauer2022cloud}
R.~Hartauer, J.~Manner, and G.~Wirtz, ``Cloud function lifecycle considerations
  for portability in function as a service.'' in \emph{Proc. CLOSER}, 2022, pp.
  133--140.

\bibitem{9305905}
D.~Taibi, J.~Spillner, and K.~Wawruch, ``Serverless computing-where are we now,
  and where are we heading?'' \emph{IEEE Software}, vol.~38, no.~1, pp. 25--31,
  2021.

\bibitem{castro2022hybrid}
P.~Castro, V.~Isahagian, V.~Muthusamy, and A.~Slominski, ``Hybrid serverless
  computing: Opportunities and challenges,'' \emph{arXiv:2208.04213}, 2022.

\bibitem{216063}
L.~Wang, M.~Li, Y.~Zhang, T.~Ristenpart, and M.~Swift, ``Peeking behind the
  curtains of serverless platforms,'' in \emph{Proc. 2018 USENIX Annu.
  Technical Conf. (USENIX ATC 18)}, Jul. 2018, pp. 133--146.

\bibitem{data_gravity}
FACTION, ``{What is Data Gravity? How it Can Influence Your Cloud Strategy},''
  \url{https://www.factioninc.com/blog/data-gravity-as-the-center-of-your-multi-cloud-universe/},
  2022.

\bibitem{serverless_framework}
{Serverless Framework}. \url{https://www.serverless.com/}. 2022.

\bibitem{microsoft_multicloud}
``{Multicloud solutions with the Serverless Framework},''
  \url{https://docs.microsoft.com/en-us/azure/architecture/example-scenario/serverless/serverless-multicloud},
  2022.

\bibitem{chasins2022sky}
S.~Chasins, A.~Cheung, N.~Crooks, A.~Ghodsi, K.~Goldberg, J.~E. Gonzalez, J.~M.
  Hellerstein, M.~I. Jordan, A.~D. Joseph, M.~W. Mahoney \emph{et~al.}, ``The
  sky above the clouds,'' \emph{arXiv:2205.07147}, 2022.

\bibitem{multicloud-trend}
P.~Iorio, ``{Multicloud trends},''
  \url{https://pablo-iorio.medium.com/multicloud-patterns-and-trends-a66ea092185e},
  2022.

\bibitem{status_multicloud}
B.~Adler, ``{Cloud Computing Trends: Flexera 2022 State of the Cloud Report},''
  \url{https://www.flexera.com/blog/cloud/cloud-computing-trends-2022-state-of-the-cloud-report},
  2022.

\bibitem{stoica2021cloud}
I.~Stoica and S.~Shenker, ``From cloud computing to sky computing,'' in
  \emph{Proc. Workshop Hot Topics in Operating Systems}, 2021, pp. 26--32.

\bibitem{lambda_price}
{AWS Lambda Pricing}. \url{https://aws.amazon.com/lambda/pricing/}. 2022.

\bibitem{gcf_price}
Cloud functions pricing. \url{https://cloud.google.com/functions/pricing}.
  2022.

\bibitem{alibaba_price}
Function compute pricing.
  \url{https://www.alibabacloud.com/product/function-compute/pricing}. 2022.

\bibitem{aws_s3}
{Amazon S3}. \url{https://aws.amazon.com/s3/}. 2022.

\bibitem{DBLP:conf/cidr/HellersteinFGSS19}
J.~M. Hellerstein, J.~M. Faleiro, J.~Gonzalez, J.~Schleier{-}Smith,
  V.~Sreekanti, A.~Tumanov, and C.~Wu, ``Serverless computing: One step
  forward, two steps back,'' in \emph{Proc. 9th Biennial Conf. Innovative Data
  Systems Research ({CIDR} 2019)}, 2019.

\bibitem{altwaijiry2021cloud}
A.~AlTwaijiry, ``Cloud computing present limitations and future trends,''
  \emph{ScienceOpen Preprints}, 2021.

\bibitem{FunctionBench}
J.~Kim and K.~Lee, ``Functionbench: A suite of workloads for serverless cloud
  function service,'' in \emph{Proc. 2019 IEEE 12th Internat. Conf. Cloud
  Computing (CLOUD 2019)}, 2019, pp. 502--504.

\bibitem{grambow2021befaas}
M.~Grambow, T.~Pfandzelter, L.~Burchard, C.~Schubert, M.~Zhao, and D.~Bermbach,
  ``Befaas: An application-centric benchmarking framework for faas platforms,''
  in \emph{Proc. 2021 IEEE Int. Conf. Cloud Engineering (IC2E 2021)}, 2021, pp.
  1--8.

\bibitem{copik2021sebs}
M.~Copik, G.~Kwasniewski, M.~Besta, M.~Podstawski, and T.~Hoefler, ``Sebs: A
  serverless benchmark suite for function-as-a-service computing,'' in
  \emph{Proc. 22nd Int. Middleware Conference}, 2021, pp. 64--78.

\bibitem{pillow}
{Pillow}. \url{https://pillow.readthedocs.io/en/stable/}. 2022.

\bibitem{jiang2021towards}
J.~Jiang, S.~Gan, Y.~Liu, F.~Wang, G.~Alonso, A.~Klimovic, A.~Singla, W.~Wu,
  and C.~Zhang, ``Towards demystifying serverless machine learning training,''
  in \emph{Proc. 2021 Internat. Conf. Management Data}, 2021, pp. 857--871.

\bibitem{carreira2018case}
J.~Carreira, P.~Fonseca, A.~Tumanov, A.~Zhang, and R.~Katz, ``A case for
  serverless machine learning,'' in \emph{Proc. Workshop Systems for ML and
  Open Source Software at NeurIPS}, 2018.

\bibitem{sanchez2021experience}
M.~S{\'a}nchez-Artigas and P.~G. Sarroca, ``Experience paper: Towards enhancing
  cost efficiency in serverless machine learning training,'' in \emph{Proc.
  22nd Internat. Middleware Conference}, 2021, pp. 210--222.

\bibitem{scikit-learn}
{scikit-learn}. \url{https://scikit-learn.org/stable/}. 2022.

\bibitem{Lithops}
J.~Sampe, P.~Garcia-Lopez, M.~Sanchez-Artigas, G.~Vernik, P.~Roca-Llaberia, and
  A.~Arjona, ``Toward multicloud access transparency in serverless computing,''
  \emph{IEEE Software}, vol.~38, no.~1, pp. 68--74, 2021.

\bibitem{vsp}
A.~F. Baarzi, G.~Kesidis, C.~Joe-Wong, and M.~Shahrad, ``On merits and
  viability of multi-cloud serverless,'' in \emph{Proc. ACM Symp. Cloud
  Computing}, 2021, pp. 600--608.

\bibitem{shedule2}
A.~Aske and X.~Zhao, ``Supporting multi-provider serverless computing on the
  edge,'' in \emph{Proc. 47th Internat. Conf. Parallel Processing Companion},
  2018, pp. 1--6.

\bibitem{shedule1}
A.~Jindal, J.~Frielinghaus, M.~Chadha, and M.~Gerndt, ``Courier: delivering
  serverless functions within heterogeneous faas deployments,'' in \emph{Proc.
  14th IEEE/ACM Internat. Conf. Utility and Cloud Computing}, 2021.

\end{thebibliography}

\end{document}